\documentclass[fleqn,usenatbib]{mnras}

\usepackage{newtxtext,newtxmath}

\usepackage[T1]{fontenc}

\DeclareRobustCommand{\VAN}[3]{#2}
\let\VANthebibliography\thebibliography
\def\thebibliography{\DeclareRobustCommand{\VAN}[3]{##3}\VANthebibliography}

\usepackage{graphicx}	
\usepackage{amsmath}	
\usepackage{multirow}
\usepackage{dcolumn}
\usepackage{xcolor}

\title[Inverse Compton scattering in high-z galaxies]{Evidence for inverse Compton scattering in high-redshift Lyman-break galaxies}

\author[I. H. Whittam et al.]{I. H. Whittam,$^{1,2}$\thanks{E-mail: imogen.whittam@physics.ox.ac.uk}
M. J. Jarvis,$^{1,2}$
Eric J. Murphy,$^{3}$
N. J. Adams,$^{4}$
R. A. A. Bowler,$^{4}$
A. Matthews,$^{5}$
\and
R. G. Varadaraj,$^1$ 
C. L. Hale,$^{1}$
I. Heywood,$^{1,6,7}$
K. Knowles,$^{6,7,8}$
L. Marchetti,$^{9,10,11}$
N. Seymour,$^{12}$
\and
F. Tabatabaei,$^{13,14,15}$
A. R. Taylor,$^{2,10,15}$
M. Vaccari,$^{10,16,9}$
A. Verma$^1$
\\
$^{1}$ Astrophysics, Department of Physics, University of Oxford, Keble Road, Oxford, OX1 3RH, UK 
\\
$^{2}$ Department of Physics and Astronomy, University of the Western Cape, Robert Sobukwe Road, 7535 Bellville, Cape Town, South Africa. 
\\
$^{3}$ National Radio Astronomy Observatory, 520 Edgemont Road, Charlottesville, VA 22903, USA\\
$^{4}$ Jodrell Bank Centre for Astrophysics, University of Manchester, Oxford Road, Manchester M13 9PL, UK\\
$^{5}$ Carnegie Observatories, 813 Santa Barbara Street, Pasadena, CA 91101, USA.\\
$^{6}$ Centre for Radio Astronomy Techniques and Technologies, Department of Physics and Electronics, Rhodes University, PO Box 94, Makhanda 6140, South Africa\\
$^{7}$ South African Radio Astronomy Observatory, 2 Fir Street, Black River Park, Observatory 7925, South Africa\\
$^{8}$ School of Mathematics, Statistics, and Computer Science, University of KwaZulu-Natal, Westville Campus, South Africa\\
$^{9}$ Department of Astronomy, University of Cape Town, 7701 Rondebosch, Cape Town, South Africa\\
$^{10}$ INAF - Istituto di Radioastronomia, via Gobetti 101, I-40129 Bologna, Italy\\
$^{11}$ The Inter-University Institute for Data Intensive Astronomy (IDIA), Department of Astronomy, University of Cape Town, 7701 Rondebosch, Cape Town, South Africa\\
$^{12}$International Centre for Radio Astronomy Research, Curtin University, GPO Box U1987, Bentley, WA 6845, Australia\\
$^{13}$ School of Astronomy, Institute for Research in Fundamental Sciences (IPM), P.O. Box 1956836613, Tehran, Iran\\
$^{14}$ Max-Planck-Institiut f\"ur Astronomy, K\"onigstul 17, 69117, Heidelberg, Germany\\
$^{15}$ Max-Planck Institut f\"ur Radioastronomie, Auf dem H\"ugel 69, 53121 Bonn, Germany\\
$^{16}$Inter-University Institute for Data Intensive Astronomy, Department of Physics and Astronomy, University of the Western Cape, 7535 Bellville, Cape Town, South Africa\\
}

\date{Accepted XXX. Received YYY; in original form ZZZ}

\pubyear{2023}

\begin{document}
\label{firstpage}
\pagerange{\pageref{firstpage}--\pageref{lastpage}}
\maketitle

\begin{abstract}
Radio continuum emission provides a unique opportunity to study star-formation unbiased by dust obscuration. However, if radio observations are to be used to accurately trace star-formation to high redshifts, it is crucial that the physical processes which affect the radio emission from star-forming galaxies are well understood. While inverse Compton (IC) losses from the cosmic microwave background (CMB) are negligible in the local universe, the rapid increase in the strength of the CMB energy density with redshift [$\sim (1+z)^4$] means that this effect  becomes increasingly important at $z\gtrsim3$.
Using a sample of $\sim200,000$ high-redshift ($3 < z < 5$) Lyman-break galaxies selected in the rest-frame ultraviolet (UV), we have stacked radio observations from the MIGHTEE survey to estimate their 1.4-GHz flux densities. We find that for a given rest-frame UV magnitude, the 1.4-GHz flux density and luminosity decrease with redshift. We compare these results to the theoretical predicted effect of energy losses due to inverse Compton scattering off the CMB, and find that the observed decrease is consistent with this explanation. We discuss other possible causes for the observed decrease in radio flux density with redshift at a given UV magnitude, such as a top-heavy initial mass function at high redshift or an evolution of the dust properties, but suggest that inverse Compton scattering is the most compelling explanation.
\end{abstract}




\section{Introduction}\label{section:intro}

Understanding how galaxies form and evolve over cosmic time is intrinsically linked to being able to measure the rate at which stars form from their molecular gas reservoirs. There are a plethora of methods for measuring the star-formation rate (SFR) in galaxies \citep[see][for reviews]{Kennicutt1998, Kennicutt2012}, traditionally by using the observed luminosity at a given wavelength that closely traces a physical process associated with star formation. For example, the most direct tracer is the ultraviolet (UV) emission from the young massive stars, however the UV spectrum is inaccessible to ground-based telescopes for relatively low-redshift ($z<0.5$) galaxies, meaning that space-based observations are required. At higher redshifts ($z>1$), the UV emission enters the visible wavelengths accessible to ground-based telescopes. However, although it is the most direct probe of the ongoing star formation in galaxies by directly measuring emission from stellar photospheres, the UV is also the most susceptible to obscuration by dust, leading to underestimates of the total star formation that may be occurring in individual systems \citep[e.g.][]{Burgarella2005, Hao2011,Buat2012,Casey2014, Heinis2014}. 

The amount of dust obscuration that occurs in star-forming galaxies is now known to be significant \citep[e.g.][]{Burgarella2013,MD2014}. Deep surveys have elucidated the nature of the star formation history of the Universe, showing that the majority of the star-formation density at $z\approx 2$ is indeed obscured by dust, with ultraviolet emission accounting for around 15-30 per cent of the total SFR density \cite[see e.g. figure 8 in][]{MD2014}. Beyond $z\sim 2$ the picture is less clear, predominantly due to the fact that most studies rely on data from the {\em Herschel Space Observatory}, which has relatively poor angular resolution and limited sensitivity at the longest wavelengths ($\gtrsim 250$\,$\muup$m) that are sensitive to the peak of the dust emission at high redshift. Although past studies with ground-based sub-mm telescopes demonstrate that there are still heavily obscured systems at the highest redshifts \citep[e.g.][]{Hughes1998, Blain2002, Geach2017}, these studies are generally not deep enough to detect the typical star-forming population that is responsible for the bulk of the SFR density at these redshifts.
Small area studies with the Atacama Large Millimetre Array (ALMA) have improved our understanding \citep[e.g.][]{Dunlop2017, Franco2018, Dudzeviciute2020}, but they are limited by small number statistics, with others based on the follow-up of UV-bright sources \citep[e.g.][]{Inami2022, Bowler2024}, which introduces more selection effects. 

Many studies have shown that the extinction due to dust can be very clumpy and vary across the galaxy \citep[e.g.][]{Bowler2022, Giminez2023, Lines2024, Harvey2024}, calling into question how one balances the relative contributions to the star-formation rate density (SFRD) from both `obscured' and `unobscured' systems. This leads to the open question of whether we can combine low-resolution far-infrared and sub-mm data with much higher resolution rest-frame UV data, where in the former we are integrating the far-infrared emission over the whole galaxy and in the latter we are potentially just sampling through the holes in the dust distribution. This calls into question the validity of spectral energy density (SED) fitting codes that determine the total SFR using the assumption of `energy balance', where all of the energy absorbed at short wavelength is balanced by the energy emitted at longer wavelengths \citep[e.g. ][]{MagPhys, CIGALE}. However, detailed studies of simulated galaxies where energy balance methods are tested on whole galaxy systems, suggest that this may not be a significant problem in determining the overall SFR of galaxies at low redshift \cite[e.g.][]{HaywardSmith2015}. At higher redshifts the picture becomes more uncertain, as star-formation becomes burstier \citep[][]{Haskell2024}, and observational samples become more biased.

Given the above, it is important that we have an independent measure of the SFR in galaxies, where uncertainties around dust obscuration and emission are minimised. An avenue for this is provided by radio continuum observations. Radio continuum emission (at $\sim 1 - 100$~GHz) from galaxies is generally due to the combination of free-free emission from H{\sc ii} regions and synchrotron radiation from electrons accelerated in magnetic fields \citep[see e.g.][ for a review]{Condon1992}. Both of these mechanisms operate in star-forming galaxies and are correlated with the star-formation rate \citep[e.g.][]{murphy2011}. Therefore, using radio continuum emission opens up a pathway to understanding the evolution of star formation in the Universe free of the effects of dust, with sufficient angular resolution to overcome confusion and the capability to cover a wide enough area to not be limited by small number statistics \citep[e.g.][]{Smolcic2008,McAlpine2013, Novak2017, Whittam2024}.

However, if radio observations are to be used to accurately trace star-formation, it is crucial that the physical processes which affect the radio emission from star-forming galaxies are well understood. Energy losses of relativistic electrons due to inverse-Compton scattering off of the CMB are negligible in the local universe, but as the strength of the CMB energy density scales as $\sim(1 + z)^4$, this effect is expected to be significant at higher redshifts. \citet{Murphy2009} uses theoretical predictions to estimate the magnitude of this effect on the observed radio signal from star-forming galaxies, and shows that it should be significant beyond $z \sim 3$. However, due to the depth of radio data required to detect radio emission from star-forming galaxies at $z \gtrsim 3$, this effect has not yet been convincingly observed.

In recent years, data taken with the Jansky Very Large Array \citep[e.g.][]{Novak2017}, MeerKAT \citep[][]{delvecchio2021,Matthews2024} and the Low-frequency Array \citep[e.g.][]{Gurkan2018, Smith2021, Cochrane2023} have been used to investigate the evolution of the star-formation rate density of the Universe, and how this may depend on galaxy properties. However, such studies are still limited by sensitivity, and these investigations of the star-formation rate traced through radio continuum observations have been limited to $z \lesssim 3$. 

An alternative approach to exploring the high-redshift faint SFG population with radio data comes through stacking approaches, where the depth of optical and near-infrared data is used to select high-redshift galaxies and then an average of the radio flux density is obtained by combining data for all these objects \citep[e.g.][]{Carilli2008, Karim2011} or by modelling the complete flux-density distribution using Bayesian techniques \citep[e.g.][]{Zwart2015, Malefahlo2022}. However, all of these previous studies have been restricted to a relatively small area ($<2$\,degrees) with varying quality of ancillary data from which to select high-redshift star-forming galaxies.

In this work, we use a sample of $3<z<5$ galaxies selected in the rest-frame UV compiled by \citet{Adams2023}, together with radio data taken with the MeerKAT telescope as part of the MeerKAT International GHz Tiered Extragalactic Exploration (MIGHTEE; \citealt{Jarvis2016}) survey to investigate the link between SFR traced by the UV continuum and the radio emission. Our sample covers the redshift range where we would expect inverse Compton scattering to be apparent from the radio emission, and the depth and areal coverage of the radio data is sufficient to perform statistical stacking on sub-samples of the Lyman-break galaxies.
This paper is laid out as follows: the data used are described in Section~\ref{section:data}, and the methods used to stack the radio data and estimate star-formation rates are outlined in Section~\ref{section:methods}. Our results are presented in Section~\ref{section:results} and discussed in Section~\ref{section:discussion}. In Section~\ref{section:conclusions} we provide our conclusions.

Throughout this paper the following values for the cosmological parameters are used: $H_0 = 70~{\rm km \, s}^{-1} ~\rm Mpc^{-1}$, $\Omegaup_{\rm M} = 0.3$ and $\Omegaup_{\Lambdaup} = 0.7$. Unless stated all magnitudes are AB magnitudes. We use the following convention for radio spectral index, $\alpha$: $S_\nu \propto \nu^{-\alpha}$, for a source with flux density $S_\nu$ and frequency $\nu$.

\section{Data used}\label{section:data}

\subsection{Sample selection}

In this work we use the sample of $\sim 230,000$ Lyman break galaxies in the redshift range  $3 < z < 5$ described in \citet{Adams2023}. The sample is selected in the rest-frame UV, and covers 10~deg$^2$ in three well-studied extra-galactic fields; XMM-LSS, COSMOS and the Extended Chandra Deep Field South (E-CDFS). 

The multi-wavelength data consist of 14 photometric bands covering $0.3-2.4~\muup$m; optical data from the Canada–France–Hawaii-Telescope Legacy Survey \citep[CFHTLS;]{Cuillandre2012}, the VST Optical Imaging of the CDFS Field \citep[VOICE;][]{Vaccari2016}, and the HyperSuprimeCam Strategic Survey Programme \citep[HSC DR2;][]{Aihara2018a,Aihara2018b,Aihara2019}. The near-infrared data in the XMM-LSS and CDFS fields are from the VISTA Deep Extragalactic Observations (VIDEO) survey \citep{Jarvis2013}, while UltraVISTA DR4 \citep[][]{McCracken2012} provides the near-infrared coverage in COSMOS. For further details we refer the reader to \citet{Bowler2021,Adams2023,Varadaraj2023}.

\begin{figure}
    \centering
    \includegraphics[width=\columnwidth]{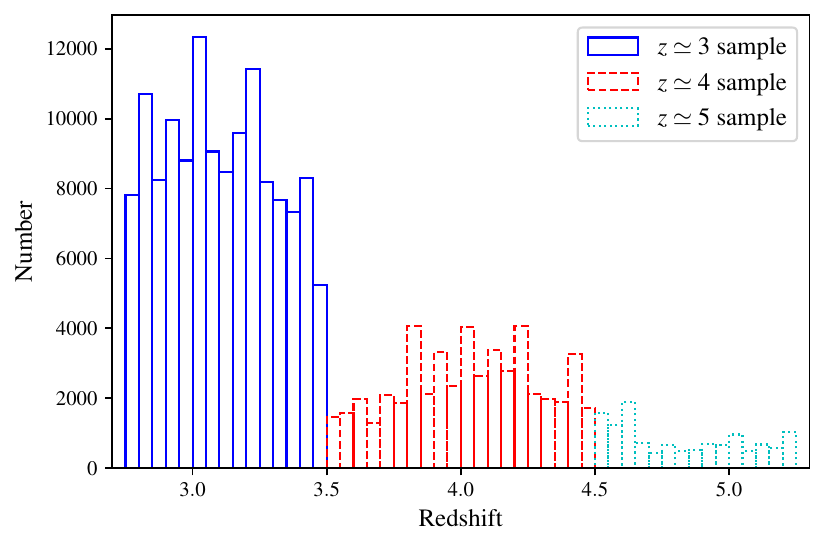}
    \caption{The redshift distribution of sources in the \citet{Adams2023} catalogue used in this work.}
    \label{fig:zdist}
\end{figure}

\begin{figure}
    \centering
    \includegraphics[width=\columnwidth]{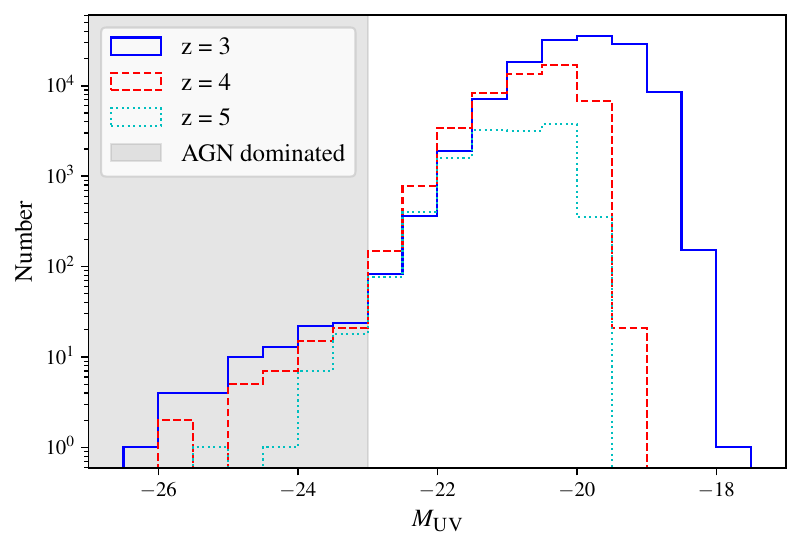}
    \caption{The absolute UV magnitude distribution of sources in the \citet{Adams2023} catalogue used in this work. Sources in the three redshift bins are shown separately. The region with $M_\textsc{uv} < -23$, where we expect the observed sources to be AGN-dominated \citep[see][]{Adams2023} is shaded in grey.}
    \label{fig:Muv_dist}
\end{figure}

Photometric redshifts are obtained using the \textsc{LePhare} SED template fitting code, full details are given in \citet{Adams2023}. 
The \citeauthor{Adams2023} sample of Lyman break galaxies used in this work consists of three different redshift samples centred on $ z \simeq 3, 4, 5$ respectively, which are selected using the following criteria summarised briefly here; we refer the reader to \citet{Adams2023} for full details. For the $z \simeq 3$ and $\simeq 4$ samples, a source is required to have a $\geq 5 \sigma$ detection in the band containing the rest-frame UV continuum emission ($r$ for $z \simeq 3$ and $i$ for $z \simeq 4$), and a photometric redshift in the range $2.75 < z < 3.5$ or $3.5 < z < 4.5$ for the two samples. The $z \simeq 5$ sample was generated by selecting sources with a $\geq 5\sigma$ detection in the HSC-z band and a photometric redshift in the range $4.5 < z < 5.2$\footnote{An upper limit of $z=5.2$ was imposed to limit contamination by brown dwarfs, see \citet{Adams2023} for further details.}. Additionally, sources in this redshift bin were required to have a $< 3\sigma$ detection in the CFHT-$u^*$ band, as the strong Lyman break exhibited by galaxies at this redshift should result in a non-detection in this band. These criteria result in a sample of $233,110$ Lyman break galaxies, with $133,200$, $49,955$ and $12,604$ galaxies in the $z \simeq 3, \simeq 4$ and $ \simeq 5$ samples respectively. The redshift distributions of the three samples are shown in Fig.~\ref{fig:zdist}, and the absolute UV magnitude distributions are shown in Fig.~\ref{fig:Muv_dist}.

\subsection{Radio Data}\label{section:radio-data}

The MIGHTEE survey \citep{Jarvis2016} is one of the Large Survey Projects currently underway with the MeerKAT radio telescope \citep{Jonas2009,Jonas2016}. The survey covers over $\sim $20~deg$^2$ in four well-studied extragalactic fields; XMM-LSS, COSMOS, ELAIS-S1 and E-CDFS. The data are primarily taken using the L-band receiver, spanning 856 - 1711~MHz. The MIGHTEE survey produces continuum \citep{Heywood2022, Hale2025}, spectral line \citep{Heywood2024} and polarisation \citep{Taylor2024} data.

In this work we use the MIGHTEE Continuum Data Release 1 (DR1) data, which is described in detail in \citet{Hale2025}. These data cover $\sim 20$~deg$^2$ in the COSMOS, XMM-LSS and E-CDFS fields, and consist of 86 individual MeerKAT tracks totalling 709.2~h across the three fields. 
Two versions of the data processed with different \citet{Briggs1995} robust parameters are released in each field; the first uses Briggs’ robustness parameter = -1.2 which down-weights the short baselines in the core, resulting in a higher resolution image, but this comes at the expense of sensitivity. The second image uses robust = 0.0, which is optimised for sensitivity but has a lower resolution. In this paper we use the higher-resolution (robust = -1.2) image, as the lower-resolution image is dominated by confusion noise, which is particularly problematic when using a stacking analysis. The high-resolution image has a circular synthesised beam full-width half maximum (FWHM) diameter of 5.2 (COSMOS), 5.0 (XMM-LSS) and 5.5 (CDFS) arcsec, and the pixel size is $1.1 \times 1.1$~arcsec in all three fields. The thermal noise in the three fields is $0.9 - 3.4~\muup$Jy/beam, however due to confusion noise the effective rms noise in the centre of the high-resolution image of each field is 2.4, 1.2 and 3.6 $\muup$Jy (COSMOS, CDFS and XMM-LSS respectively).

Due to the wide bandwidth of the MeerKAT L-band receiver (856 - 1711~MHz) and the varying response of the primary beam with frequency (along with other factors such as flagging of the raw data), the effective frequency of the MIGHTEE DR1 data varies across the image \citep[see][for details]{Heywood2022,Hale2025}. We therefore make use of the effective frequency map released with the DR1 images to scale the flux density of each source to 1.4 GHz, this is discussed further in Section~\ref{section:methods}.

\section{Methods}\label{section:methods}

\subsection{Stacking the radio data}\label{section:stacking}

\begin{figure}
    \centering
    \includegraphics[width=0.95\columnwidth]{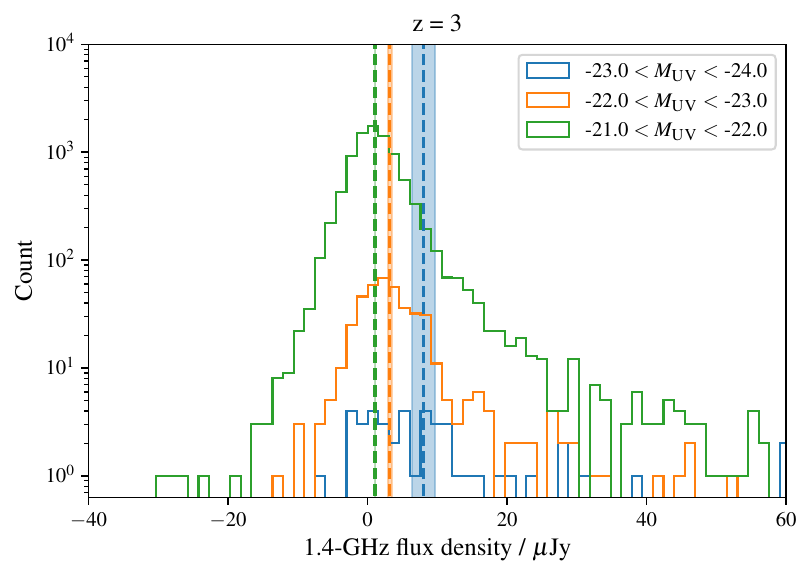}
    \includegraphics[width=0.95\columnwidth]{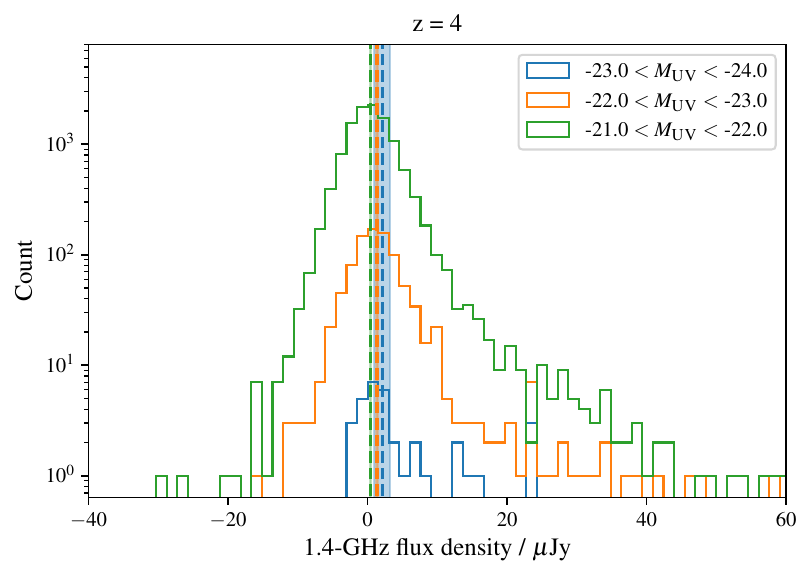}
    \includegraphics[width=0.95\columnwidth]{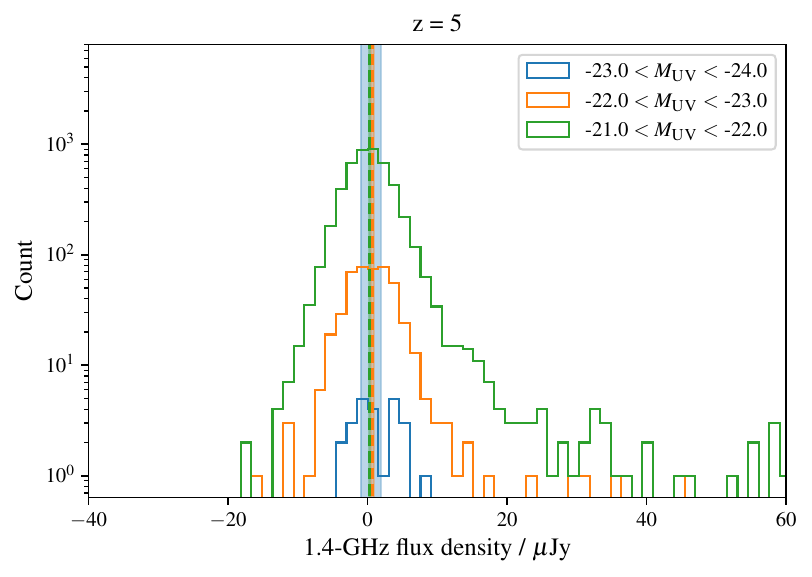}
    \caption{Distributions of corrected pixel flux densities in each $M\textsc{uv}$ bin (each pixel value is corrected for local background level and scaled to 1.4 GHz). The vertical dashed lines show the median flux density in each $M\textsc{uv}$ bin, and the shaded coloured regions illustrate the uncertainties on these median flux densities, estimated by bootstrap re-sampling. The different panels show the three redshift bins. The $M\textsc{uv}$ bins are the same as those used in Fig.~\ref{fig:IC}.}
    \label{fig:flux-dist}
\end{figure}

In order to estimate the radio flux densities of the high-redshift galaxies in the \citet{Adams2023} sample,
 we measure the flux density of the pixel in the radio image at the position of that object. The synthesised beam full-width half maximum is $\sim$5 arcsec (it varies slightly between fields), so we expect all of the galaxies in our sample to be unresolved, given the largest UV size is $\sim4$\,kpc ($\sim0.5$~arcsec) at these redshifts \citep{Varadaraj2024}. This means that the central pixel provides a good estimate of the total flux density of each source.
We then measure radio pixel values at 500 randomly generated positions in a box $1.8 \times 1.8$~arcmin around the object (excluding the central region where the source lies) and find the median random pixel value (i.e.\ an estimate of the local background level). This median random flux density is then subtracted from the pixel flux density value for the object to give an estimate of the radio flux density of the object. Each flux density is then scaled to 1.4~GHz assuming a spectral index of $\alpha = 0.7$\footnote{See section~\ref{section:discussion-alpha} for a discussion on the effect of choice of spectral index on the results.}, taking into account the effective frequency at each position in the radio image (as the frequency varies slightly across the radio images, see Section~\ref{section:radio-data}). The galaxies in each of the three redshift samples are split into bins according to their rest-frame UV-magnitude ($M_\textsc{uv}$), and the median stacked radio flux density in each bin is calculated. The uncertainties in the median values are estimated using bootstrap re-sampling. The distributions of the corrected radio pixel flux densities and the medians in each bin are shown in Fig.~\ref{fig:flux-dist}, and the median flux densities are shown as a function of $M_\textsc{uv}$ in Fig.~\ref{fig:medianS14}. 

To estimate the stacked radio luminosity, the corrected radio flux density of each object is converted to a radio luminosity, using the photometric redshift of each individual object and applying a k-correction assuming a radio spectral index of $\alpha = 0.7$. The median radio luminosity and uncertainty (using bootstrap re-sampling) is then calculated in each $M_\textsc{uv}$ / redshift bin, and the resulting distributions are shown in Fig.~\ref{fig:medianL}. The median stacked flux densities, luminosities and associated uncertainties are shown in Table~\ref{tab:s14_L14}, along with the number of sources in each bin.

\begin{table*}
\scriptsize{
\centering
\begin{tabular}{c|ccr|ccr|ccr}
\hline
\multirow{2}{*}{$M_\textsc{uv}$ Bin} & \multicolumn{3}{c|}{z = 3} & \multicolumn{3}{c|}{z = 4} & \multicolumn{3}{c}{z = 5} \\
 & $S_{1.4}$ / $\mu$Jy beam$^{-1}$ & $L_{1.4~}$ / W Hz$^{-1}$ & $N$
& $S_{1.4}$ / $\mu$Jy beam$^{-1}$ & $L_{1.4}$ / W Hz$^{-1}$ & $N$
& $S_{1.4}$ / $\mu$Jy beam$^{-1}$ & $L_{1.4}$ / W Hz$^{-1}$ & $N$\\
\hline
-20.0 $< M_\textsc{uv} <$ -20.5 & 0.20 $\pm$ 0.02 & (1.10 $\pm$ 0.11)$\times$10$^{22}$ & 31711 & -- & -- & -- & -- & -- & -- \\
-20.5 $< M_\textsc{uv} <$ -21.0 & 0.44 $\pm$ 0.03 & (2.45 $\pm$ 0.15)$\times$10$^{22}$ & 18142 & 0.05 $\pm$ 0.03 & (4.86 $\pm$ 3.33)$\times$10$^{21}$ & 13342 & -- & -- & -- \\
-21.0 $< M_\textsc{uv} <$ -21.5 & 0.92 $\pm$ 0.04 & (5.15 $\pm$ 0.23)$\times$10$^{22}$ & 7175 & 0.34 $\pm$ 0.04 & (3.35 $\pm$ 0.33)$\times$10$^{22}$ & 8338 & 0.13 $\pm$ 0.07 & (1.91 $\pm$ 1.02)$\times$10$^{22}$ & 3232 \\
-21.5 $< M_\textsc{uv} <$ -22.0 & 1.57 $\pm$ 0.10 & (8.57 $\pm$ 0.45)$\times$10$^{22}$ & 1906 & 0.61 $\pm$ 0.07 & (6.03 $\pm$ 0.73)$\times$10$^{22}$ & 3448 & 0.43 $\pm$ 0.09 & (6.16 $\pm$ 1.32)$\times$10$^{22}$ & 1606 \\
-22.0 $< M_\textsc{uv} <$ -22.5 & 2.93 $\pm$ 0.33 & (1.64 $\pm$ 0.17)$\times$10$^{23}$ & 359 & 1.18 $\pm$ 0.13 & (1.15 $\pm$ 0.13)$\times$10$^{23}$ & 771 & 0.75 $\pm$ 0.21 & (9.87 $\pm$ 2.89)$\times$10$^{22}$ & 398 \\
-22.5 $< M_\textsc{uv} <$ -23.0 & 4.08 $\pm$ 0.77 & (2.39 $\pm$ 0.41)$\times$10$^{23}$ & 82 & 3.08 $\pm$ 0.47 & (3.15 $\pm$ 0.44)$\times$10$^{23}$ & 147 & 0.69 $\pm$ 0.51 & (1.04 $\pm$ 0.75)$\times$10$^{23}$ & 76 \\
-23.0 $< M_\textsc{uv} <$ -24.0 & 8.03 $\pm$ 1.76 & (4.20 $\pm$ 1.13)$\times$10$^{23}$ & 46 & 2.09 $\pm$ 1.10 & (2.14 $\pm$ 1.03)$\times$10$^{23}$ & 36 & 0.43 $\pm$ 1.40 & (7.03 $\pm$ 20.48)$\times$10$^{22}$ & 25 \\
-24.0 $< M_\textsc{uv} <$ -25.0 & 7.85 $\pm$ 2.65 & (4.05 $\pm$ 1.78)$\times$10$^{23}$ & 23 & 6.50 $\pm$ 2.89 & (5.59 $\pm$ 2.30)$\times$10$^{23}$ & 12 & -- & -- & -- \\
-25.0 $< M_\textsc{uv} <$ -26.0 & 16.22 $\pm$ 5.02 & (8.84 $\pm$ 2.60)$\times$10$^{23}$ & 8 & -- & -- & -- & -- & -- & -- \\
\hline
\end{tabular}}
\caption{Median 1.4 GHz flux density ($S_{1.4}$), luminosity ($L_{1.4}$) and number of sources ($N$) for each $M_\textsc{uv}$ and redshift bin. The flux densities listed here are the median corrected pixel values (in $\muup$Jy/beam); assuming the sources are unresolved, these pixel values correspond to the total flux densities (in $\muup$Jy), see text for details. Uncertainties on the medians are estimated using bootstrap resampling.}
\label{tab:s14_L14}
\end{table*}


\citet{Carilli2008} stacked the 1.4-GHz radio flux densities of a considerably smaller sample of Lyman break galaxies at $z \sim 3, 4,$ and 5 using VLA data in the COSMOS field with a median rms of $\sim 15~\muup$Jy beam$^{-1}$. They found that the $z \sim 3$ sample, containing 6457 galaxies, had a median stacked 1.4-GHz flux density of $0.90 \pm 0.21~\muup$Jy. A random subsample of 6457 galaxies drawn from our $z \sim 3$ sample has a stacked median flux density of $0.29 \pm 0.05~\muup$Jy, which is $\sim 3 \sigma$ lower than the \citeauthor{Carilli2008} result. This difference is not surprising as the parent datasets are very different, and the stacked radio flux density is strongly dependent on host galaxy properties, such as $M_\textsc{uv}$ (e.g. see Fig.~\ref{fig:medianS14}). The \citeauthor{Carilli2008} $z=3$ detection is consistent with our $z = 3$ sample with $M_\textsc{uv} > -21$.
The \citet{Carilli2008} $z \sim 4$ and 5 samples, with 1447 and 614 galaxies respectively, were undetected in their radio stacks. The upper limits from their work are higher than the median stacked flux densities in our $z \sim 4$ and 5 samples, and therefore consistent.

\subsection{The far-infrared radio correlation}\label{section:SFRs}

The far-infrared - radio correlation (FIRC) can be used to estimate the star-formation rate (SFR) of a galaxy from its radio luminosity with the underlying assumption that the total infrared luminosity of a galaxy provides a robust measurement of its SFR. The FIRC can be quantified by the parameter $q_\text{IR}$, which is defined as the logarithmic ratio of the infrared and radio luminosities:
\begin{equation}
    q_\text{IR} = \text{log}_{10} \frac{L_\text{IR}~[\text{W}] \, / \, 3.75 \times 10^{12} ~[\text{Hz}]} {L_{1.4~\text{GHz}}~[\text{W / Hz}]}\label{eqn:qir}
\end{equation}
where $L_\text{IR}$ is the total infrared luminosity between 8 - 1000 $\muup$m. This is divided by the central frequency of $3.75 \times 10^{12} ~\text{Hz}$ (80 $\muup$m) so that $q_\text{IR}$ is a dimensionless quantity. The radio SFR can then be estimated from the radio luminosity as follows:
\begin{equation}
    \textrm{SFR} [M_\odot / \textrm{yr}] = f_\textrm{IMF}^{-1}  \, 10^{-24} \, 10^{q_\textrm{IR}} L_{1.4~\textrm{GHz}} [\textrm{W / Hz}]
\end{equation}
where $f_\textrm{IMF}$ is a factor accounting for the IMF ($f_\textrm{IMF} = 1$ for a \citeauthor{Chabrier2003} IMF, used here).

In order to compare to theoretical predictions and other work, and to investigate the implications of our results of star-formation rate estimates, we estimate $q_\textrm{IR}$ values for the galaxies in our sample in the following way. The UV star-formation rate is calculated from the UV magnitude using the scaling from \citet{MD2014}, and assuming a \citet{Chabrier2003} initial mass function (IMF). As we are not able to constrain the possible attenuation values, we make no attempt to correct these values for dust absorption, meaning that these values are likely to be underestimates of the SFR. The potential effects of dust on these results are discussed in Section~\ref{section:dust}. We then scale from SFR to infrared luminosity ($L_\textrm{IR}$) using the updated \citet{Kennicutt2012} calibration, which is based on the work by \citet{murphy2011,Hao2011}, as follows:  
\begin{equation}
    L_\textrm{IR} [L_\odot] =  8.6 \times 10^{9} \, f_\textrm{IMF}^{-1}  \, \textrm{SFR} [M_\odot / \textrm{yr}].\label{eqn:LIRSFR}
\end{equation}
The $q_\textrm{IR}$ values in each $M_\textsc{uv}$ bin are then estimated from the median stacked radio luminosities using equation~\ref{eqn:qir}.
As mentioned previously, the $M_\textsc{uv}$ values are uncorrected for dust meaning that they may be underestimated, causing the derived $q_\textrm{IR}$ values to also be underestimates.

\section{Results}\label{section:results}

\begin{figure}
    \centering
    \includegraphics[width=\columnwidth]{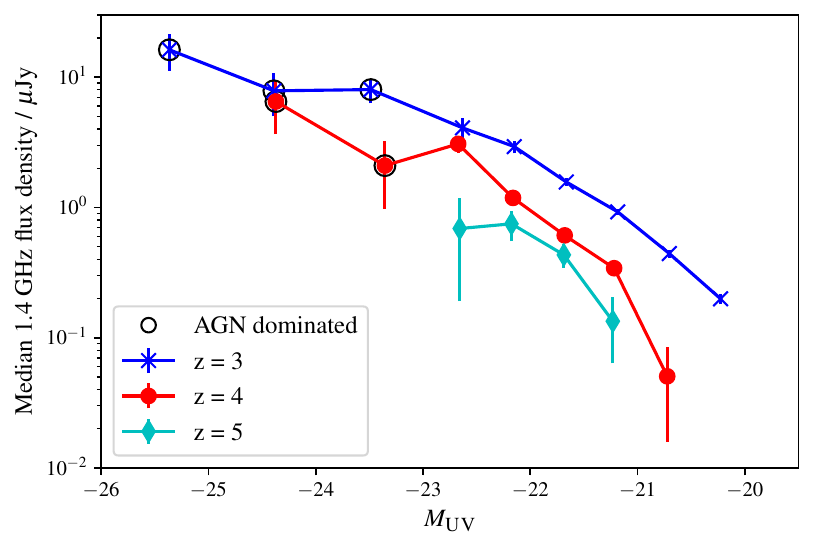}
    \caption{Median 1.4-GHz stacked flux density, corrected for the background, in $M_\textsc{uv}$ bins. Bins which are dominated by AGN are marked by black circles. The three redshift bins are shown separately. Error bars plotted are the uncertainty on the median estimated from bootstrap re-sampling. Points are plotted at the median $M_\textsc{uv}$ value in each bin.}
    \label{fig:medianS14}
\end{figure}

\begin{figure}
    \centering
    \includegraphics[width=\columnwidth]{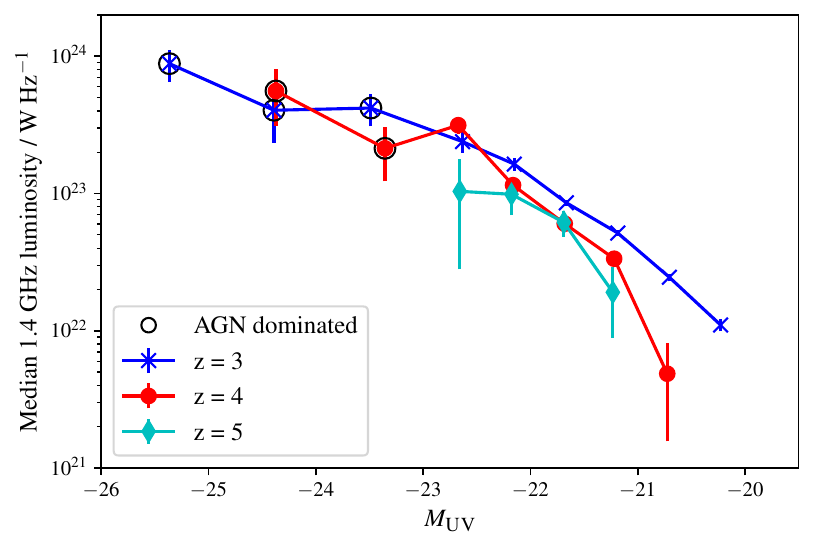}
    \caption{Median 1.4-GHz radio luminosity in $M_\textsc{uv}$ bins for the three fields, with the three redshift bins shown separately. Bins which are dominated by AGN are marked by black circles. Error bars plotted are the uncertainties on the median estimated from bootstrap re-sampling. Points are plotted at the median $M_\textsc{uv}$ in each bin.}
    \label{fig:medianL}
\end{figure}


The median stacked radio flux density and luminosity as a function of absolute UV-magnitude are shown in Figs.~\ref{fig:medianS14} and \ref{fig:medianL}, respectively. It is clear that for a given UV magnitude, both the radio flux density and luminosity decrease as redshift increases. We defer discussion of this to Section~\ref{section:discussion}.

\begin{figure}
    \centering
    \includegraphics[width=\columnwidth]{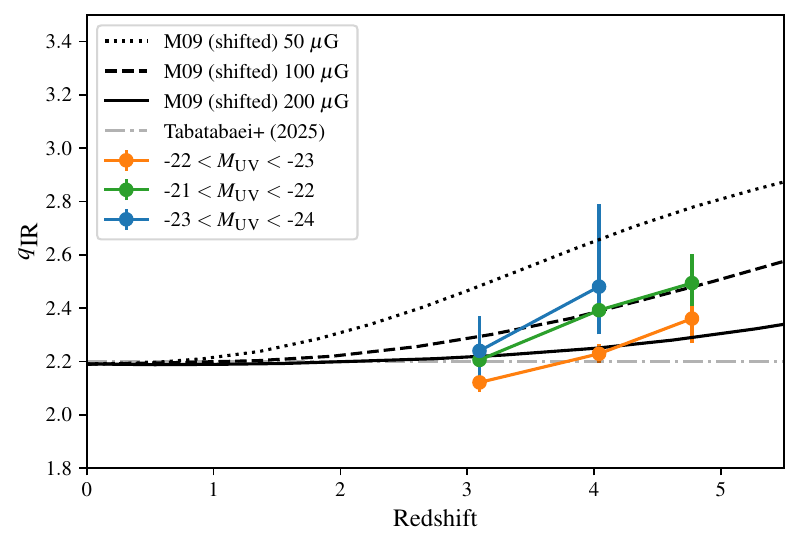}
    \includegraphics[width=\columnwidth]{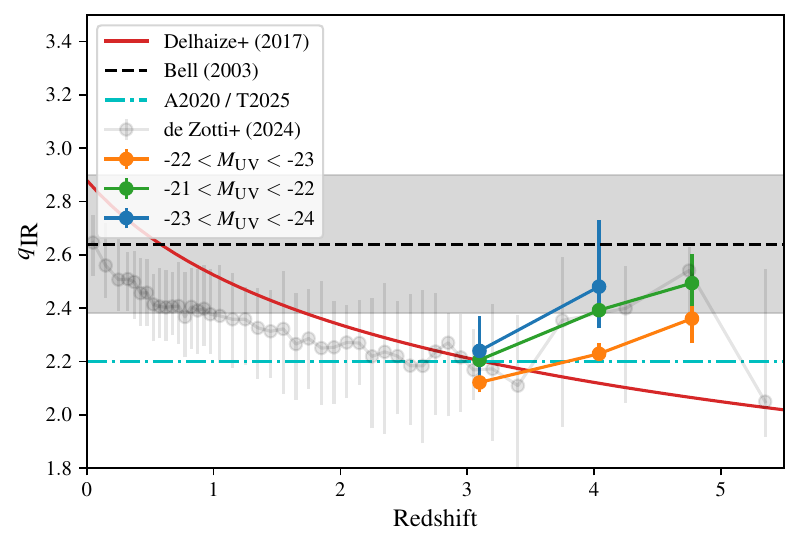}
    \caption{$q_\textrm{IR}$ as a function of redshift. The coloured points show the stacked results from this work, in the same $M_\textsc{uv}$ and redshift bins shown in Fig.~\ref{fig:IC}. We note that no correction for dust obscuration has been applied, so these are likely to be underestimates, and that the highest $M_\textsc{uv}$ bin ($-23 < M_\textsc{uv} < -24$) is likely to dominated by AGN emission. In the \textbf{top panel}, the black lines show the predicted relations from \citet{Murphy2009}, which take into account the effect of IC scattering (adapted from figure 5 of \citet{Murphy2009}). These are shifted by -0.45 in $q_\textrm{IR}$ from the original \citet{Murphy2009} relations to line up with the \citet{Tabatabaei2025} observational result at 1.5 < z < 3.5 (shown by the grey dot-dashed line, which also serves to illustrate the no-IC scattering case). In the \textbf{bottom panel} several $q_\textrm{IR}$ relations from observational studies are also shown; the  black dashed line shows the \citet{Bell2003} relation, with the grey shaded region illustrating the associated spread (0.26). The solid red line shows the power-law fit from \citet{Delhaize2017}, and the cyan dash-dot line shows the \citet{Algera2020,Tabatabaei2025} relation. The grey points are the median values from Table 1 of \citet{dezotti2024}.} 
    \label{fig:qir}
\end{figure}

Fig.~\ref{fig:qir} shows the $q_\textrm{IR}$ values as a function of redshift, in three $M_\textsc{uv}$ bins. In the top panel, our results are compared to the theoretical predictions of \citet{Murphy2009}, which take into account the effects of Inverse-Compton scattering from the CMB. The \citeauthor{Murphy2009} relations are shifted by -0.45 in $q_\textrm{IR}$ to line up with the recent \citet{Algera2020} and \citet{Tabatabaei2025} observational results at z < 3. Our results clearly show an increase in $q_\textrm{IR}$ with redshift, with a slope consistent with the predicted effects of IC scattering. This is discussed further in Section~\ref{section:discussion-IC}. 

In the bottom panel of Fig.~\ref{fig:qir} our results are compared to several other observational studies; the first is the \citet{Bell2003} value $q_\textrm{IR} = 2.64 \pm 0.02$, which was derived at low redshift. The second is the redshift-dependent relationship from \citet{Delhaize2017}; $q_\text{IR}=(2.88 \pm 0.03)(1 + z)^{-0.19\pm0.01}$. The physical processes responsible for this observed redshift evolution are open to debate, as recent work has suggested that a dependence on stellar mass \citep[e.g.][]{delvecchio2021,Smith2021} or an evolution of the spectral index may be responsible for the observed trend. The third relation shown is from recent work by \citet{Tabatabaei2025}, which uses significantly deeper radio data and finds $q_\textrm{IR} = 2.20 \pm 0.01$ over $1.5 < z < 3.5$, with no evidence for redshift evolution. This value is consistent with previous work by \citet{Algera2020}, who found $q_\textrm{IR} = 2.20 \pm 0.03$, with no evidence for redshift evolution, for a sample of submillimeter galaxies over a similar redshift range.  
The \citet{Delhaize2017} sample has recently been re-analysed by \citet{dezotti2024} with the addition of the MIGHTEE Early Science data now available in the field \citep{Heywood2022}, the results from their work are shown as grey points.

Again, we note that since the $M_\textsc{uv}$ values which are used to estimate the $q_\textrm{IR}$ values are uncorrected for dust attenuation and should therefore be considered lower limits, the $q_\textrm{IR}$ values should be considered as lower limits. As such, while our results are consistent with the \citet{Delhaize2017}, \citet{Algera2020} and \citet{Tabatabaei2025} results at $z \sim 3$, at $z > 3.5$ we find $q_\textrm{IR}$ values larger than these previous works, which may become more discrepant when accounting for the fact that these are essentially lower limits. Our results show remarkable agreement with \citet{dezotti2024}, particularly since their work estimates $L_\textrm{IR}$ directly from \emph{Herschel} data, rather than scaling from the UV emission as done in our work. This is discussed further in Section~\ref{section:dust}. The implications of these results for radio SFR estimates are discussed in Section~\ref{section:discussion-SFR}. 

\section{Discussion}\label{section:discussion}

In Section~\ref{section:results} we found that for a given rest-frame UV magnitude, the stacked rest-frame 1.4-GHz flux density decreases with redshift. We discuss possible reasons for this result in this section, and discuss the implications for radio SFR estimations in Section~\ref{section:discussion-SFR}.

\subsection{Inverse Compton scattering off the CMB}\label{section:discussion-IC}

\begin{figure}
    \centering
    \includegraphics[width=\columnwidth]{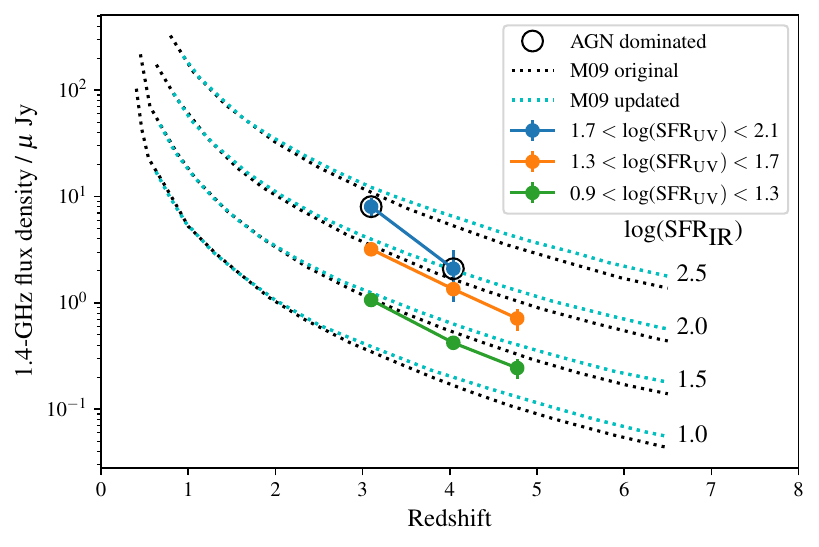}
    \caption{Observed-frame 1.4-GHz flux densities as a function of redshift. The coloured points show the observed values from this work, for three UV SFR bins. These bins correspond the the following $M_\textsc{uv}$ ranges: $-24 < M_\textsc{uv} < -23$ (blue), $-23 < M_\textsc{uv} < -22$ (orange) and $-22 < M_\textsc{uv} < -21$ (green). We note that the brightest $M\textsc{uv}$ bin (blue) is dominated by AGN. The $z \simeq 5$ point is not shown for the $1.7 < \textrm{(log(SFR}_\textrm{UV}) < 2.1$ bin as the uncertainties on this value are very large ($0.43 \pm 1.40 \muup$Jy). The points are plotted at the median redshift in each bin. The dotted black lines show the theoretical predicted relationships from \citet{Murphy2009}, taking into account energy losses of electrons due to IC scattering off of the CMB, assuming an internal magnetic field strength of 100~$\muup$G. The cyan dotted lines show an updated version of the \citet{Murphy2009} prediction, see text for details. The SFRs are given in $\textrm{M}_\odot \, \textrm{yr}^{-1}$.}
    \label{fig:IC}
\end{figure}

Energy losses of relativistic electrons to CMB photons via inverse-Compton scattering are negligible in the local universe, but are expected to become increasingly significant at $z \gtrsim 3$, reducing the observed radio flux density originating from synchrotron emission. Recent work by \citet{dezotti2024} investigated the redshift-dependence of the far-infrared - radio correlation by combining the \citet{Delhaize2017} sample with MIGHTEE Early Science data in the COSMOS field \citep{Heywood2022}. They found a hint of an upturn in $q_\textrm{IR}$ at $z>3.5$, which they note is consistent with the predicted decrease in synchrotron emission due to IC scattering off the CMB, but they were not able to probe high enough redshifts to confirm this result.

A relative decrease in the radio flux density with redshift, over and above that expected from cosmological dimming, is clear in the work presented here.  In Fig.~\ref{fig:IC} we show this in three UV SFR bins, where the UV SFR is estimated from the rest-frame UV magnitude as described in Section~\ref{section:SFRs}, and three redshift bins. 
\cite{Murphy2009} used theoretical predictions to estimate the observed radio flux density as a function of redshift, taking into account the effect of IC scattering off the CMB. These theoretical predictions are shown as the black dotted lines in Fig.~\ref{fig:IC}, for a range of different infrared-derived star-formation rates. The original \citeauthor{Murphy2009} predictions assumed a constant radiation field energy density ($U_\textrm{rad}$) for the galaxies, but in Fig.~\ref{fig:IC} we also show an updated version (cyan dotted lines), where $U_\textrm{rad}$ is allowed to scale with $U_\textrm{B}$, the magnetic field energy density of the galaxy \citep[][]{murphy2015}. 
The predicted flux densities in \citet{Murphy2009} are provided in infrared luminosity bins, but to ease comparison with our sample in Fig.~\ref{fig:IC} we have converted these to infrared SFRs, using the scaling in equation~\ref{eqn:LIRSFR}.

The decrease in the radio flux density seen in the observed data is in general agreement with the decrease due to IC scattering predicted by \citet{Murphy2009}. For the higher SFRs (log$_\textrm{10}$(SFR) $> 1.3 \textrm{M }_\odot \, \textrm{yr}^{-1}$), the observed decrease in flux density with redshift appears to be steeper than predicted. However, we expect the UV emission in this bin, which corresponds to UV magnitudes of $-24 < M_\textsc{uv} < -23$, to be dominated by AGN emission, as discussed further in Section~\ref{section:AGN}. The decrease in flux density with redshift is slightly steeper than predicted for the remaining two UV SFR bins; more data are required to determine if this steepening is significant, and whether it is indicative of a change in the magnetic field strength, or other effects. For example, the skewed redshift distributions in each bin and the lack of UV dust attenuation corrections could be playing a role. For a given SFR and redshift, the observed flux density is also higher than predicted. The updated \citet[][]{murphy2015} predictions, where $U_\textrm{rad}$ is allowed to scale with $U_\textrm{B}$, are a better match to our observations.

The comparison between our observational results and the \citet{Murphy2009} predicted effect of inverse-Compton scattering is visualised in a different way in Fig.~\ref{fig:qir}, which shows $q_\textrm{IR}$ as a function of redshift. In the top panel, the results from this work are compared to the \citeauthor{Murphy2009} predictions, which show how IC-scattering is predicted to cause the radio flux density to decrease, and therefore $q_\textrm{IR}$ increase, with increasing redshift. In this plot the \citeauthor{Murphy2009} relations are shifted by -0.45 in $q_\textrm{IR}$ to line up with the recent \citet{Tabatabaei2025} $q_\textrm{IR}$ result at $z < 3$. The redshift-invariant \citeauthor{Tabatabaei2025} relation is also shown in the figure, and serves to illustrate the no-IC losses case. The data clearly show an increase in $q_\textrm{IR}$ with redshift, with a slope consistent with the predicted effect of IC scattering. 

The strength of the IC signal depends on the internal magnetic field strength of each galaxy. \citet{Murphy2009} predicts the magnitude of this effect using two different assumptions for the internal magnetic field strength; 10~$\muup$G and 100~$\muup$G. The predictions assuming 100~$\muup$G are a better match to our observations, and are the ones shown in Fig.~\ref{fig:IC}. 
This is consistent with recent work by \citet{Tabatabaei2025} who investigated the magnetic fields strength of star-forming galaxies detected in MIGHTEE in the redshift range $1.5 < z < 3.5$. They find that magnetic field strength increases with increasing redshift such that $B = (55 \pm 7) \times (1 + z)^{(0.7 \pm 0.1)} \muup \textrm{G}$. Evaluating this relation at the mean redshift value for our three redshift samples gives magnetic field strengths of 147, 170 and 189~$\muup$G for the $ z = 3, 4,$ and 5 samples respectively.

Overall, this provides evidence that the decrease in flux density observed with redshift in our stacked sample is consistent with the predictions for inverse Compton scattering of cosmic ray electrons off of the CMB, and provides compelling observational evidence for this effect. Other possible causes of the observed decrease are discussed in the remainder of the section.

This result differs from what was shown in figure 6 of \citet{Murphy2009}, where they collected detections from the literature and found that the observed $q_\textrm{IR}$ values fall below the expectations due to IC losses from the CMB. This is because the observations available at the time only contain the most luminous objects in the sky, and do not include the more typical galaxies at high redshift. With the significantly deeper data available here, we are able to extract the average properties of much fainter populations of galaxies, and investigate the typical galaxy population. This work shows that when the general galaxy population is considered, the observations are consistent with the predicted losses due to IC scattering.

\subsection{Effect of dust}\label{section:dust}

As discussed in Section~\ref{section:intro}, UV magnitudes, and therefore the UV-derived SFRs, are sensitive to dust obscuration. As we do not have enough information to constrain the dust content of the galaxies in this sample, we do not attempt to correct the UV magnitudes for dust attenuation. The true UV magnitudes of the galaxies in our sample are therefore likely to be higher than the values used here, as some of the emission will be absorbed by dust in the galaxy.

\begin{figure}
    \centering
    \includegraphics[width=\columnwidth]{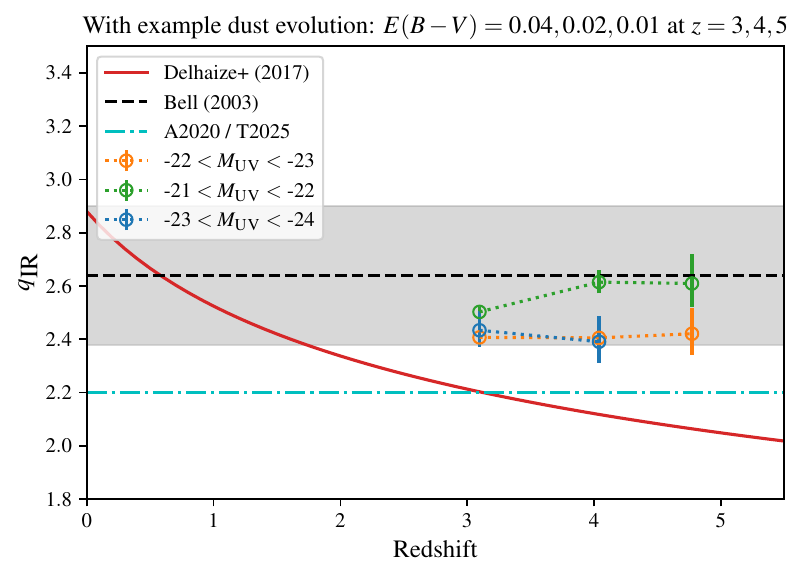}
    \caption{Investigating the potential effects of dust evolution on these results. Here we reproduce Fig.~\ref{fig:qir} ($q_\textrm{IR}$ as a function of redshift) with one possible dust evolution, fine-tuned to remove the observed drop in radio luminosity relative to UV luminosity (and therefore increase in $q_\textrm{IR}$) with redshift; $E(B-V) = 0.04, 0.02, 0.01$ at redshifts 3, 4, 5 respectively. The coloured points show the stacked results from this work with a correction for the example evolution of dust attenuation applied, in the same $M_\textsc{uv}$ and redshift bins shown in Fig.~\ref{fig:IC}. We note that the highest $M_\textsc{uv}$ bin ($-23 < M_\textsc{uv} < -24$) is likely to dominated by AGN emission. Several $q_\textrm{IR}$ relations from observational studies are also shown; the  black dashed line shows the \citet{Bell2003} relation, with the grey shaded region illustrating the associated spread (0.26). The solid red line shows the power-law fit from \citet{Delhaize2017}, and the cyan dash-dot line shows the \citet{Algera2020,Tabatabaei2025} relation. }
    \label{fig:test_EBV}
\end{figure}

Although radio emission is unaffected by dust, we have binned the galaxies in our sample according to their uncorrected rest-frame UV magnitude when stacking the radio data. This means that if the level of dust attenuation varies with redshift, it is possible that this could be responsible for the observed relative decrease in radio flux density (compared to the UV emission), with redshift, rather than IC scattering. To test the possible effects of dust evolution, we investigate the several different possible prescriptions for the evolution of dust attenuation on these results, to see if any could reproduce the observed decrease in radio luminosity for a given UV magnitude. We find that that if the dust properties of this sample of Lyman-break galaxies evolve such that the mean $E(B-V)$ values at $z = 3,4,5$ are $E(B-V) = 0.04, 0.02, 0.01$ respectively, then the observed decrease in radio luminosity relative to UV luminosity can be accounted for. This is demonstrated in Fig.~\ref{fig:test_EBV}, where Fig.~\ref{fig:qir} is reproduced, with the rest-frame $M_\textsc{uv}$ values corrected by this example $E(B-V)$ distribution. As these $E(B-V)$ values are chosen to remove the observed decrease in radio luminosity relative to UV luminosity with redshift, by design $q_\textrm{IR}$ shows no evolution with redshift.

However, our results without dust correction show a remarkably good agreement with  \citet{dezotti2024}, as seen in Fig.~\ref{fig:qir}. This is significant because \citeauthor{dezotti2024} uses infrared data from \emph{Herschel} to calculate the $L_\textrm{IR}$ values directly, rather than scaling from the UV emission as done here, so they will be affected by dust in the opposite way to our results (i.e. if dust attenuation decreased with redshift, this would cause the UV luminosities and therefore our $q_\textrm{IR}$ estimates to increase, but would cause the infrared luminosities and therefore the \citeauthor{dezotti2024} $q_\textrm{IR}$ values to decrease). Therefore, the fact that the two results are in agreement suggests that dust is not the cause of the increase in $q_\textrm{IR}$ observed.

Additionally, since the galaxies are all selected via the Lyman break technique in the rest-frame UV, they are expected to have reasonably similar dust properties, so there is unlikely to be a significant change in their dust content with redshift even is there is a general change in dust properties for the wider galaxy populations at these redshifts. Additionally, observational evidence from the VANDELS spectroscopic survey \citep{Cullen2018} suggests that dust attenuation does not evolve over the redshift range $0 < z < 5$. This is supported by the mild evolution seen in the colour-magnitude relation over $z = 3$--$5$ \citep[e.g.][]{Bouwens2014,Morales2025}. More recently, \citet{Bowler2024} used ALMA data to investigate the properties of Lyman-break galaxies and found little evolution in the average dust properties between $4 < z < 8$.  Therefore, we suggest that it is unlikely that dust attenuation is responsible for the decrease in radio luminosity, relative to the UV luminosity, with redshift observed here. 

\subsection{Contribution of AGN}\label{section:AGN}

If AGN activity is present, this could potentially contribute to the emission in either the UV or radio wavelengths, or both. \citet{Bowler2021,Adams2023} estimated the fraction of AGN in the sample used here by parameterising the UV luminosity function to include functional forms for both galaxies and AGN. They found that AGN dominate (i.e. contribute $ \gtrsim 90$ per cent of) the $z = 3$ sample at $M_\textsc{uv} \lesssim -23.0$, the $z = 4$ sample at $M_\textsc{uv} \lesssim -23.5$ and the $z = 5$ sample at $M_\textsc{uv} \lesssim -24$.  Galaxies dominate the three samples at $M_\textsc{uv} \gtrsim -22, -22.5$ and $-22.5$ respectively, with a transition between the two populations in between. The three UV SFR bins shown in Fig.~\ref{fig:IC} correspond to the following $M_\textsc{uv}$ values; $-24 < M_\textsc{uv} < -23, -23 < M_\textsc{uv} < -22, -22 < M_\textsc{uv} < -21$. So while we expect the brightest bin to be dominated by AGN, the contribution of AGN to the faintest bin should be negligible. Given that the decrease in radio flux density with luminosity is seen in all three UV-magnitude bins shown in Fig.~\ref{fig:IC}, it is unlikely that AGN emission in the UV is contributing significantly to this result. Moreover, the low radio luminosities of the galaxies in this sample (see Fig.~\ref{fig:medianL}) imply that there is unlikely to be significant AGN contribution to the radio emission \citep[see e.g.][]{Mauch2007,Smolcic2017,Novak2017,Malefahlo2022}. 

The effect of AGN activity in the UV-bright sources is evident when looking at the radio luminosity distributions in Fig.~\ref{fig:medianL}; there is a clear turnover in the distributions at $M_\textsc{uv} < -23$. This is most likely due to AGN emission contributing to the UV flux, while not contributing significantly to the radio flux density,
causing the UV-derived SFRs to be over-estimated when compared to the radio SFRs.

\subsection{Size evolution}\label{section:sizes}

A decrease in radio luminosity with increasing redshift could also occur if galaxies at higher redshift are systematically smaller than those at lower redshift. This could mean that cosmic rays diffuse out of the galaxy on timescales shorter than those required to maintain the FIRC \citep{Yun2001,Carilli2008, murphy2008}. \citet{Varadaraj2024} studied the sizes of a sub-sample of the Lyman break galaxies used in this work using data from the JWST Public Release IMaging for Extragalactic Research (PRIMER) survey \citep{Dunlop2021}. They found no significant evolution in the median size of galaxies between $3 < z < 5$, therefore size evolution is unlikely to be responsible for the decrease in radio luminosity with redshift observed in this work.

\subsection{Evolution in the Initial Mass Function}

\begin{figure}
    \centering
    \includegraphics[width=\columnwidth]{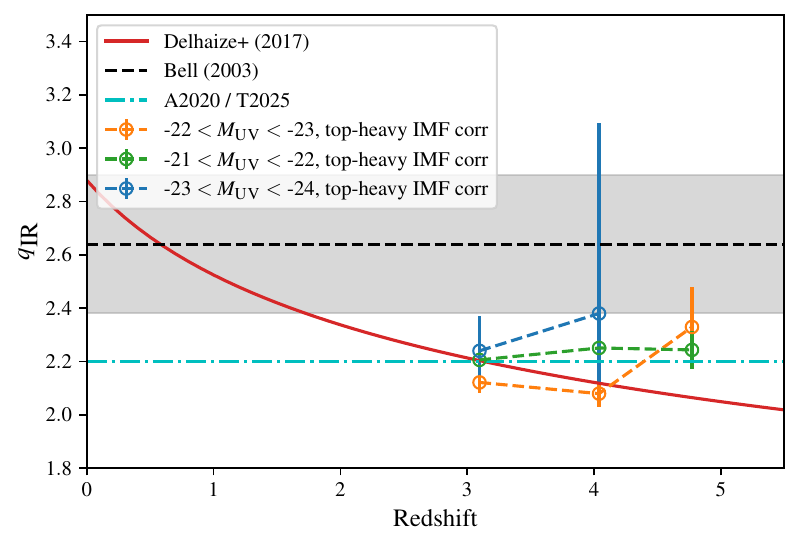}
    \caption{Investigating the possible effects of an evolving IMF. Here we reproduce Fig.~\ref{fig:qir}, $q_\textrm{IR}$ as a function of redshift, correcting for the effect of one possible evolution of the IMF. The coloured points show the stacked results from this work, with the $z = 4$ and 5 samples corrected for the effect of a top-heavy IMF with slope -2.0 at $>25 \, M_\odot$ (see text for details). We note that no correction for dust obscuration has been applied, so these are likely to be underestimates, and that the highest $M_\textsc{uv}$ bin ($-23 < M_\textsc{uv} < -24$) is likely to dominated by AGN emission. Several $q_\textrm{IR}$ relations from observational studies are also shown; the  black dashed line shows the \citet{Bell2003} relation, with the grey shaded region illustrating the associated spread (0.26). The solid red line shows the power-law fit from \citet{Delhaize2017}, and the cyan dash-dot line shows the \citet{Algera2020,Tabatabaei2025} relation. }
    \label{fig:IMF}
\end{figure}

Another possibility is an evolution of the initial mass function (IMF). Radio emission at $\sim$GHz frequencies is produced by massive stars ($\gtrsim 8 M_\odot$) which end their lives in core-collapse supernovae. However very massive stars, with $M \gtrsim 25 M_\odot$, typically end their lives via direct collapse or exotic supernovae, and may not produce radio synchrotron emission from supernovae remnants (see e.g.\ \citealt{Gehrz1983,Smartt2009}). UV emission, on the other hand, is produced by all stars with stellar masses $\gtrsim 3 M_\odot$. This means that if the IMF has changed between $3<z< 5$, then the ratio between the UV and radio emission caused by star-formation will also have changed. For example, if the IMF was more top-heavy at higher redshift, then this would mean that there were a larger fraction of stars with masses $> 25 M_\odot$, which would lead to more UV emission relative to the radio emission. Indeed, a top-heavy IMF has been suggested as a possible explanation for the abundance of very bright sub-millimetre galaxies at high redshift \citep[e.g.][]{Baugh2005}, and recent work by  \citet{Zhang2018} and \citet{Cameron2024} find tentative evidence for a top-heavy IMF at redshifts $z \sim 2 - 3$ and $z = 6$ respectively.

To investigate the possible impact of an evolving IMF on our results, we estimate the increase in UV luminosity relative to radio luminosity that would be caused by the high-mass end of the IMF at $M > 25$\,M$_\odot$ changing from a \citet{Chabrier2003} IMF to a more top-heavy IMF. We find that if the IMF changes such that the power-law slope is -2.0 at $M > 25$\,M$_\odot$, as opposed to a \citet{Chabrier2003} IMF with slope = -2.3, this would cause the ratio of the UV to radio luminosities to increase by a factor of 1.27, a relative change in rest-frame UV magnitude of -0.26. To give an indication of the impact this effect could have on our results, we `corrected' for the effect by reducing the UV magnitudes of all galaxies in our $z = 4$ and 5 samples by this amount; in Fig.~\ref{fig:IMF} we reproduce the results of Fig.~\ref{fig:qir} with this correction. This shows that if the IMF changed from a Chabrier IMF to a more top-heavy IMF with slope$=-2.0$ at $M > 25$\,M$_\odot$ in the time between redshifts 3 and 4, then this could potentially explain the reduction in radio flux relative to UV flux we observe.

\citet{murphy-imf2011} also investigate the effect of different IMFs on the radio and UV luminosities. Taking into account the changes at the low-mass end, they find that changing from a Kroupa IMF to a top-heavy IMF with a power-law slope of -1.5 would change the ratio of the radio to UV luminosities by a factor of 1.35, which would also approximately account for the observed decrease in radio luminosity for a given UV luminosity, if this change occurred between redshifts 3 and 4.

It is therefore possible to construct plausible evolutionary scenarios of the IMF which could potentially explain the observed decrease in radio luminosity for a given UV luminosity with redshift. Note that it would also be possible to combine this effect with the effect of dust evolution, for example a small evolution in the dust attenuation with redshift combined with a change in the IMF could be constructed such that it would explain the signal observed here. However, these explanations require a level of fine-tuning of the dust and/or IMF evolution over a relatively short period of cosmic time, together with IC scattering from the CMB \emph{not} behaving as predicted. We therefore conclude that IC scattering from the CMB is the most likely cause of the observed reduction in radio flux density with redshift.

\subsection{Spectral indices}\label{section:discussion-alpha}

To estimate the 1.4-GHz flux densities and luminosities in this paper we have assumed a radio spectral index of $\alpha = 0.7$. The choice of spectral index makes very little difference to the calculated flux densities, as this is only used to scale between the effective frequency ($\sim 1.2~\textrm{GHz}$) and 1.4~GHz, so cannot be responsible for the trend seen in Fig.~\ref{fig:IC}. As there is evidence for a somewhat flatter spectral index for faint star-forming galaxies at high-redshift \citep[e.g.][]{Murphy2017}, we investigated the effect of assuming $\alpha=0.6$ instead of 0.7 on our results. This would shift the luminosity distributions slightly, but does not affect any of the overall trends and therefore has no impact the conclusions of this paper. At higher redshifts we are sampling a higher rest-frame frequency, so the non-thermal contribution to the total flux density may be higher, causing the spectral shape to be flatter at higher redshifts. We therefore also tested using a simple evolving spectral index ($\alpha = 0.7, 0.6, 0.5$ at $z = 3, 4, 5$ respectively). This again has a negligible effect on the flux densities, and makes the decrease in luminosity with redshift more pronounced.

\subsection{Effect on star-formation rate estimates}\label{section:discussion-SFR}

Radio continuum observations provide a unique opportunity to estimate SFRs which is unbiased by dust. However, if these estimates are to be accurate at high-redshift, it is vital that all the physical processes which affect radio emission from star-forming galaxies are accounted for. We have shown in Fig.~\ref{fig:qir} that the \citet{Delhaize2017}, \citet{Algera2020} and \citet{Tabatabaei2025} $q_\textrm{IR}$ relations are underestimates at $z > 3.5$, leading to an underestimate for the SFR. As discussed earlier in this section, the most likely cause of the observed increase in $q_\textrm{IR}$ at $z > 3$ is inverse-Compton scattering from the CMB.
This illustrates the importance of quantifying the effect of inverse-Compton scattering if radio data are to be used as a reliable measure of star-formation rate at $z > 3.5$.

\section{Conclusions}\label{section:conclusions}

Using a sample of $\sim200,000$ high-redshift ($3 < z < 5$) galaxies selected in the rest-frame UV, we have stacked the MIGHTEE radio data to estimate the 1.4-GHz flux densities of the galaxies. We find that for a given UV magnitude, the 1.4-GHz flux density and luminosity decrease with redshift. We compare these results to the theoretical predicted effect of inverse-Compton scattering of relativistic electrons off CMB photons from \citet{Murphy2009}, and find that the observed decrease is consistent with what is predicted. 

We discuss a number of different possible causes for the observed decrease in radio luminosity relative to UV luminosity with redshift, such as a top-heavy IMF at high redshift or evolution in the dust properties. Although it is possible to construct evolutions of the IMF or dust attenuation (or a combination of the two) which could account for the observed trend, these would all require a level of fine-tuning of the IMF and/or dust evolution, together with the relatively well-understood physics of IC scattering from the CMB \emph{not} behaving as predicted. This does not seem particularly likely, particularly for galaxy samples selected in exactly the same way, over a very short period of cosmic time. We therefore suggest that IC scattering is the most compelling explanation, given that our results are consistent with the theoretical predictions of this effect, providing observational evidence of the detection of IC scattering from the CMB in high-redshift star-forming galaxies.

In the near future it should be possible to confirm this result using $2-3$~GHz data in these fields from the S-band component of the MIGHTEE survey \citep{Jarvis2016}, which is currently in progress. This will allow us to constrain the shape of the radio spectra of the galaxies in this sample, and test whether or not this is consistent with the predicted effect of IC scattering as a function of frequency.

Radio emission has long been touted as a promising method to trace star-formation, as it is unbiased by dust. However, if radio emission is to be used to accurately estimate star-formation rates at $z \gtrsim 3$, the reduction in synchrotron emission due to IC scattering off the CMB needs to be accounted for. Our results show that the \citet{Delhaize2017}, \citet{Algera2020} and \citet{Tabatabaei2025} $q_\textrm{IR}$ relations are underestimates compared to this sample at $z > 3.5$. Further work will allow us to pin down the factors which affect the magnitude of this signal, such as the internal magnetic field strength, and therefore the correction needed to be applied radio SFRs.

\section*{Acknowledgements}

We thank the anonymous referee for their helpful comments which improved the manuscript. We thank Dan Smith and Aayush Saxena for useful discussions. IHW, MJJ and CLH acknowledge support from the Oxford Hintze Centre for Astrophysical Surveys which is funded through generous support from the
Hintze Family Charitable Foundation. 
MJJ and IH acknowledge support of the STFC consolidated grant [ST/S000488/1] and [ST/W000903/1] and from a UKRI Frontiers Research Grant [EP/X026639/1]. 
NJA acknowledges support from the ERC Advanced Investigator Grant EPOCHS (788113).
RB acknowledges support from an STFC Ernest Rutherford Fellowship [grant number ST/T003596/1].
KK acknowledges funding support from the South African Radio Astronomy Observatory (Grant UID 97930).
LM acknowledges financial support from the South African Department of Science and Innovation’s National Research Foundation under the ISARP RADIOMAP Joint Research Scheme (DSI-NRF Grant Number 150551) and the CPRR Projects (DSI-NRF Grant Number SRUG2204254729).
MV acknowledges financial support from the Inter-University Institute for Data Intensive Astronomy (IDIA), a partnership of the University of Cape Town, the University of Pretoria and the University of the Western Cape, and from the South African Department of Science and Innovation's National Research Foundation under the ISARP RADIOMAP Joint Research Scheme (DSI-NRF Grant Number 150551) and the CPRR HIPPO Project (DSI-NRF Grant Number SRUG22031677).
The MeerKAT telescope is operated by the South African Radio Astronomy Observatory, which is a facility of the National Research Foundation, an agency of the Department of Science and Innovation. This work made use of Astropy:\footnote{http://www.astropy.org} a community-developed core Python package and an ecosystem of tools and resources for astronomy \citep{astropy:2013, astropy:2018, astropy:2022}.

\section*{Data Availability}

 The MIGHTEE continuum DR1 data used in this work are released with \citet{Hale2025}. The galaxy catalogues are released with \citet{Adams2023}.



\bibliographystyle{mnras}
\bibliography{lybreak_stacking} 





\bsp	
\label{lastpage}
\end{document}